\documentclass[11pt]{article}
\usepackage[left=1in, right=1in, top=1in, bottom=1.5in]{geometry}

\usepackage{amssymb, latexsym, times, graphicx, amsmath, float}
\usepackage{hyperref}
\usepackage{xcolor}
\hypersetup{
    colorlinks=true,
    linkcolor=blue,
    filecolor=magenta,      
    urlcolor=teal,
    citecolor=violet
}
\usepackage{tocbibind}

\usepackage{amstext}
\usepackage{amsthm}

\usepackage{tabu}

\usepackage{setspace}
\newtheorem{remark}{Remark}[section]

\newtheorem{proposition}{Proposition}[section]

\newtheorem{definition}{Definition}[section]
\numberwithin{equation}{section}
\newtheorem{example}{Example}[section]
\begin{document}
\title{Traveling Wave Solutions For Newton's Equations of Celestial Mechanics: Kepler's Problem}
\author{Harry Gingold and Jocelyn Quaintance\\
 {\small West Virginia University, Morgantown, WV, USA; gingold@math.wvu.edu}\\
{\small University of Pennsylvania, Philadelphia, PA, USA; jocelynq@seas.upenn.edu}}
\maketitle

\abstract{This article produces wave equations and constructs traveling
wave solutions that are intimately related to Newton's
equations of celestial mechanics. The traveling wave solutions are
expressed in ``closed form'' in
terms of elementary functions. They are specialized to the $2$-body and
the relative $2$-body problem. The traveling wave solutions disclose the shape
and position of wave fronts emanating from collisions by determining
the location of the singularities of the traveling wave solutions.}

\medskip
\textbf{Keywords:} Newton's celestial mechanics equations (NCME); $2$-body
problem; relative $2$-body problem; d'Alembert's wave equation;
companion wave equation; wave fronts; gravitational waves.

\medskip
\textbf{AMS Classification:} Primary 70F15; Secondary 85A04.

\section{Introduction}
The purpose of this article is to produce wave equations and to construct
traveling wave solutions that are intimately related to Newton's
equations of celestial mechanics. The traveling wave solutions are
expressed in ``closed form'' in
terms of elementary functions.  These wave equations contain all the terms of
Newton's equations of celestial mechanics as well as
all analogous terms present in d'Alembert's wave equation.
This article guarantees the existence of these traveling wave solutions
for the $2$-body and
the relative $2$-body problem.  The  $2$-body problem models the collisions of
binary stars, the dynamics of which produce gravitational waves.
Compare e.g. with \cite{AlNADA,BanikhovaCelestial,celestialMechanTaff,FizpatricCelestialMech,
KemphaviGRAVITWAVES,MaggioreGRAVWAVES,MillerGravitationalWaves,Moser,
NASAGRAVITWAVes,SCHmitzRelativityGravitatWaves}.
The theory of ODEs that is related to our methodology can be found
in \cite{HILLEODESCOMPLEX-1,JORDAN=000026SMITH-1,
KRANTZRealAnalytic,L. Perko-1,P.F.Hsieh=000026Y. Sibuya-2}.
A sample of treatises on wave propagation can be found in \cite{Brekhovski,Witham}.
Mathematical treatises of celestial mechanics are \cite{POLARDCELESTIAL-1,WintnerAurel}.
We apologize for not being able to mention many other excellent sources. 

\medskip
The current approach does not utilize Einstein\textquoteright s field
equations of general relativity. Compare e.g. with \cite{SCHmitzRelativityGravitatWaves}.
Nevertheless, our wave equations, which are solely based on Newton's
equations of celestial mechanics, relate to general relativity in the
following manner. Einstein, Infel, and Hoffman \cite{EinsteinInfeld}
derived a set of celestial mechanics equations that are based on general
relativity. Their equations turn out to be small perturbations of
Newton's equations of celestial mechanics with
perturbation parameter $c^{^{-1}}$, where $c$ is the speed of
light. 

\medskip
Our work produces, qualitatively, the shape and the position of the
wave front emanating from a collision of a binary star.  Parameters are available that
can be adjusted to fit experimental data with our qualitative model. 
It is noteworthy that Zeldovitch \cite{Zeldovith1969} proposed
a framework that qualitatively models gravitational instability. 
Our work is analogous to \cite{Zeldovith1969}
and allows generalized coordinates of position for the wave equation that we produce. 

\medskip
The order of topics in this article is as follows. In Section 2 we show how to extend the second order
vector differential equation $d^2y/dt^2 = f(y)$, with $y, f(y)\in\mathbb{R}^{pq}$, into a companion wave equation
\[
\frac{\partial^2\Psi}{\partial t^2} -\lambda^2\Delta_{\tilde{r}}\Psi= f(\Psi)
\]
where $\Psi:\mathbb{R}^{\ell m+1}\rightarrow\mathbb{R}^{pq}$;
see (\ref{generalizedwaveequ}).
In Section 3 we analyze the companion wave equation when
 $\Psi(\tilde{r},t) = \Psi(w)$, where $w:= v\cdot \tilde{r} - \mu t + c$
with $\mu, c \in \mathbb{R}$ and $v\in \mathbb{R}^{ml}$;
see (\ref{Specialcasgenwaveeq}).  
In Section 4 we specialize the companion wave equation to both the $2$-body problem and the relative $2$-body problem.
We generate closed form solutions to the following two companions wave equation systems, namely
\begin{align*}
\left[\mu^2 - \lambda_1^2\|v\|^2\right]\frac{d^2\Psi_1}{dw^2} &= \frac{Gm_2(\Psi_2- \Psi_1)}{\|\Psi_2-\Psi_1\|^3}\notag\\
\left[\mu^2 - \lambda_2^2\|v\|^2\right]\frac{d^2\Psi_2}{dw^2} &= \frac{Gm_1(\Psi_1- \Psi_2)}{\|\Psi_2-\Psi_1\|^3},
\end{align*}
and 
\[
[\mu^2 - \lambda_{12}^2\|v\|^2]\frac{d^2\Psi_{12}}{dw^2} = -\frac{G(m_1+m_2)\Psi_{12}}{\|\Psi_{12}\|^3};
\]
see Propositions \ref{Newton2bodyprop} and \ref{2bodyrelprop} respectively.  
Then in Section 5 we provide a physical interpretation for Proposition \ref{2bodyrelprop} which describes the shape 
of the gravitational wave fronts associated
with the collision of two binary masses.

\section{Generalized Wave Equation Associated with a Second Order Vectorial ODE}
Suppose we have a second order differential equation
\begin{equation}\label{starteq}
\frac{d^2y}{dt^2} = f(y),\qquad y, f(y)\in \mathbb{R}^{pq}.
\end{equation}
where $f:\mathbb{R}^{pq}\rightarrow \mathbb{R}^{pq}$ is continuously differentiable over some open region 
$U\subseteq\mathbb{R}^{pq}$ and the (real) independent variable of (\ref{starteq}) is $t$.
In (\ref{starteq}) we set
\begin{align}\label{yexpression}
y&:= \left[ y_1(t)\,\,
y_2(t)\,\,
\hdots\,\,
y_q(t)
\right]^T\in \mathbb{R}^{pq},\notag\\
y_j(t)&:=
\left[
y_{1j}(t),
y_{2j}(t),
\hdots,
y_{pj}(t)
\right]^T\in\mathbb{R}^p
,\,\,\,\text{for all $1\leq j\leq q$,}
\end{align}
where $T$ denotes the transpose.
In a similar manner, we also set
\begin{align}\label{fyexpression}
f(y)&:= \left[
f_1(y)\,\,
f_2(y)\,\,
\hdots\,\,
f_q(y)
\right]^T\in \mathbb{R}^{pq},\notag\\
f_j(y):&=
\left[
f_{1j}(y),
f_{2j}(y),
\hdots,
f_{pj}(y)
\right]\in\mathbb{R}^p
,\,\,\,\text{for all $1\leq j\leq q$.}
\end{align}
Our first goal is to associate with (\ref{starteq}) a modified wave equation specific to the vector field 
$f(y)$.
To that end we replace the independent variable $t$ with an $(\ell m+1)$-tuple $(\tilde{r}_1,\tilde{r}_2,\dots, \tilde r_{m},t)$, where
\begin{equation}\label{extendedrdef}
\tilde{r}:=\left[\tilde{r}_1\,\,\tilde{r}_2\,\,\hdots\,\,\tilde r_{m}\right]^T\in\mathbb{R}^{\ell m},\quad 
\tilde{r}_j:=\left[\tilde{x}_{1j},\tilde{x}_{2j},\dots,\tilde{x}_{\ell j}\right]^T\in\mathbb{R}^\ell,\,\,\,\text{for all $1\leq j\leq m$.}
\end{equation}
We define a mapping $\Psi:\mathbb{R}^{\ell m +1}\rightarrow\mathbb{R}^{pq}$, where
\begin{align}\label{Psidef}
\Psi(\tilde{r},t) &:=
\left[
\Psi_1(\tilde{r},t)\,\,
\Psi_2(\tilde{r},t)\,\,
\hdots\,\,
\Psi_q(\tilde{r},t)
\right]^T\in \mathbb{R}^{pq},\notag\\
\Psi_j(\tilde{r},t):&=
\left[
\Psi_{1j}(\tilde{r},t),
\Psi_{2j}(\tilde{r},t),
\hdots,
\Psi_{pj}(\tilde{r},t)
\right]^T\in\mathbb{R}^p
,\,\,\,\text{for all $1\leq j\leq q$,}
\end{align}
and replace each instance of $y_{kj}(t)$ with $\Psi_{kj}(\tilde{r},t)$.
We assume that each $\Psi_{kj}(\tilde{r},t)$, where $1\leq k\leq p$ and $1\leq j\leq q$, 
is a real valued function which is continuously differentiable over some open region $U_1\subseteq\mathbb{R}^{\ell m+1}$.
Recall that for $1\leq j\leq q$ and $1\leq k\leq p$, 
\begin{equation}\label{Laplacianop}
\Delta_{\tilde{r}}\Psi_{kj}(\tilde{r},t):=\sum_{s=1}^m\sum_{i=1}^{\ell}
\frac{\partial^2\Psi_{kj}}{\partial \tilde{x}_{is}^2}.
\end{equation} 
We replace the left side of (\ref{starteq}) with the d'Alembert wave operator and formally obtain
\begin{equation}\label{generalizedwaveequ}
\frac{\partial^2\Psi}{\partial t^2} -\lambda^2\Delta_{\tilde{r}}\Psi= f(\Psi),
\end{equation}
where $\lambda^2$ is the following $pq\times pq$ matrix
\begin{equation}\label{deltadef}
\lambda^2:=
\left[
\begin{array}{ccccc}
\lambda_{1,1}^2 & 0 & 0 & \hdots & 0\vspace{0.15cm}\\
0 & \lambda_{2,2}^2 & 0 & \hdots & 0\vspace{0.15cm}\\
0 & 0 & \lambda_{3,3}^2 & \hdots & 0\\
\vdots &\vdots & \vdots & \ddots & \vdots\\
0 & 0 & 0 & \hdots & \lambda^2_{pq,pq}
\end{array}
\right],\qquad\text{$\lambda_{j,j}\in \mathbb{R}/\{0\}$ whenever $1\leq j\leq pq$,}
\end{equation}
 $\Delta_{\tilde{r}}\Psi$ is the following $pq\times 1$ matrix
\begin{align}\label{DeltadLapvector}
\Delta_{\tilde{r}}\Psi&:=
\left[
\Delta\Psi_1\,\,
\Delta\Psi_2\,\,
\hdots\,\,
\Delta\Psi_q
\right]^T\in \mathbb{R}^{pq},\notag\\
\Delta\Psi_j&:=
\left[
\Delta\Psi_{1j}(\tilde{r},t),
\Delta\Psi_{2j}(\tilde{r},t),
\hdots,
\Delta\Psi_{pj}(\tilde{r},t)
\right]^T\in\mathbb{R}^p
,\,\,\,\text{for all $1\leq j\leq q$,}
\end{align}
and $f(\Psi)$ is the following $pq\times 1$ matrix,
\begin{align}\label{fPsiexpression}
f(\Psi)&:= \left[
f_1(\Psi)\,\,
f_2(\Psi)\,\,
\hdots\,\,
f_q(\Psi)
\right]^T\in \mathbb{R}^{pq},\notag\\
f_j(\Psi)&:=
\left[
f_{1j}(\Psi),
f_{2j}(\Psi),
\hdots,
f_{pj}(\Psi)
\right]^T\in\mathbb{R}^p
,\,\,\,\text{for all $1\leq j\leq q$.}
\end{align}
Equation (\ref{generalizedwaveequ}) is called the {\it companion wave equation} to (\ref{starteq}).

\begin{remark}
Throughout this article we consider (\ref{generalizedwaveequ}) and (\ref{starteq}) as related but distinct
entities.
\end{remark}

\begin{example}\label{firstexample}
Let $p = 3$, $q =1$, $\ell = 3$, and $m = 1$.  Then (\ref{extendedrdef}) becomes
\[
\tilde{r} := [\tilde{r}_1],
\qquad\text{with}\,\,
\tilde{r}_1:= [\tilde{x}_{11}, \tilde{x}_{21},\tilde{x}_{31}]^T,
\]
and (\ref{Psidef})
 becomes
\[
\Psi(\tilde{r},t) := [\Psi_1(\tilde{r},t)],\qquad\text{with}\,\,
\Psi_1(\tilde{r},t) :=[\Psi_{11}(\tilde{r},t),
\Psi_{21}(\tilde{r},t),
\Psi_{31}(\tilde{r},t)]^T.
\]
In this instance, for $1\leq k\leq 3$, Equation (\ref{Laplacianop}) becomes 
\[
\Delta_{\tilde{r}}\Psi_{k1}(\tilde{r},t) = \frac{\partial^2\Psi_{k1}}{\partial\tilde{x}_{11}^2} + 
\frac{\partial^2\Psi_{k1}}{\partial\tilde{x}_{21}^2} +\frac{\partial^2\Psi_{k1}}{\partial\tilde{x}_{31}^2}, 
\]
which means that (\ref{generalizedwaveequ}) 
with 
\[
\lambda^2 := 
\left[
\begin{array}{ccc}
\lambda_1^2 & 0 & 0 \\
0 & \lambda_1^2 & 0 \\
0 & 0 &\lambda_1^2 \\
\end{array}
\right]
\]
 becomes
\[
\left[
\begin{array}{c}
\frac{\partial^2\Psi_{11}}{\partial t^2}
\vspace{0.3cm}\\
\frac{\partial^2\Psi_{21}}{\partial t^2}
\vspace{0.3cm}\\
\frac{\partial^2\Psi_{31}}{\partial t^2}
\end{array}\right]
-
\left[
\begin{array}{c}
\lambda_1^2\Delta_{\tilde{r}}\Psi_{11}(\tilde{r},t)
\vspace{0.3cm}\\
\lambda_1^2\Delta_{\tilde{r}}\Psi_{21}(\tilde{r},t)\vspace{0.3cm}\\
\lambda_1^2\Delta_{\tilde{r}}\Psi_{31}(\tilde{r},t)\vspace{0.3cm}\\
\end{array}
\right]
 =
 \left[
\begin{array}{c}
f_{11}(\Psi_1,\Psi_2)
\vspace{0.3cm}\\
f_{21}(\Psi_1,\Psi_2)
\vspace{0.3cm}\\
f_{31}(\Psi_1,\Psi_2)
\end{array}
\right] :=
f(\Psi_1,\Psi_2).
\]
\end{example}

\begin{example}\label{firstAexample}
Let $p = 3$, $q =2$, $\ell = 3$, and $m = 1$.  Then $\tilde{r}$ is defined as in Example \ref{firstexample},
and (\ref{Psidef})
 becomes
\[
\Psi(\tilde{r},t) := 
\left[
\begin{array}{c}
\Psi_1(\tilde{r},t)\\
\Psi_2(\tilde{r},t)
\end{array}
\right],
\,\,
\Psi_1(\tilde{r},t) :=
\left[
\begin{array}{c}
\Psi_{11}(\tilde{r},t)\\
\Psi_{21}(\tilde{r},t)\\
\Psi_{31}(\tilde{r},t)
\end{array}
\right],
\,\,
\Psi_2(\tilde{r},t) :=
\left[
\begin{array}{c}
\Psi_{12}(\tilde{r},t)\\
\Psi_{22}(\tilde{r},t)\\
\Psi_{32}(\tilde{r},t)
\end{array}
\right].
\]
In this instance, for $1\leq k\leq 3$ and $1\leq j\leq 2$, Equation (\ref{Laplacianop}) becomes 
\[
\Delta_{\tilde{r}}\Psi_{kj}(\tilde{r},t) = \frac{\partial^2\Psi_{kj}}{\partial\tilde{x}_{11}^2} + 
\frac{\partial^2\Psi_{kj}}{\partial\tilde{x}_{21}^2} +\frac{\partial^2\Psi_{kj}}{\partial\tilde{x}_{31}^2}, 
\]
which means that (\ref{generalizedwaveequ}) 
with 
\[
\lambda^2 := 
\left[
\begin{array}{cccccc}
\lambda_1^2 & 0 & 0 & 0 & 0 & 0\\
0 & \lambda_1^2 & 0  & 0 & 0 & 0\\
0 & 0 &\lambda_1^2 & 0 & 0 & 0\\
0 & 0 & 0 & \lambda_2^2 & 0 & 0\\
0 & 0 & 0 & 0 & \lambda_2^2 & 0\\
0 & 0 & 0 & 0 & 0 & \lambda_2^2
\end{array}
\right]
\]
 becomes
\[
\left[
\begin{array}{c}
\frac{\partial^2\Psi_{11}}{\partial t^2}
\vspace{0.3cm}\\
\frac{\partial^2\Psi_{21}}{\partial t^2}
\vspace{0.3cm}\\
\frac{\partial^2\Psi_{31}}{\partial t^2}
\vspace{0.3cm}\\
\frac{\partial^2\Psi_{12}}{\partial t^2}
\vspace{0.3cm}\\
\frac{\partial^2\Psi_{22}}{\partial t^2}\vspace{0.3cm}\\
\frac{\partial^2\Psi_{32}}{\partial t^2}
\end{array}\right]
-
\left[
\begin{array}{c}
\lambda_1^2\Delta_{\tilde{r}}\Psi_{11}(\tilde{r},t)
\vspace{0.3cm}\\
\lambda_1^2\Delta_{\tilde{r}}\Psi_{21}(\tilde{r},t)\vspace{0.3cm}\\
\lambda_1^2\Delta_{\tilde{r}}\Psi_{31}(\tilde{r},t)\vspace{0.3cm}\\
\lambda_2^2\Delta_{\tilde{r}}\Psi_{12}(\tilde{r},t)\vspace{0.3cm}\\
\lambda_2^2\Delta_{\tilde{r}}\Psi_{22}(\tilde{r},t)\vspace{0.3cm}\\
\lambda_2^2\Delta_{\tilde{r}}\Psi_{32}(\tilde{r},t)
\end{array}
\right]
 =
 \left[
\begin{array}{c}
f_{11}(\Psi_1,\Psi_2)
\vspace{0.3cm}\\
f_{21}(\Psi_1,\Psi_2)
\vspace{0.3cm}\\
f_{31}(\Psi_1,\Psi_2)
\vspace{0.3cm}\\
f_{12}(\Psi_1,\Psi_2)\vspace{0.3cm}\\
f_{22}(\Psi_1,\Psi_2)\vspace{0.3cm}\\
f_{32}(\Psi_1,\Psi_2)
\end{array}
\right] :=
f(\Psi_1,\Psi_2).
\]
\end{example}

\section{The Companion Wave Equation}
We now look at a special instance of (\ref{generalizedwaveequ}), namely the case when
\begin{equation}\label{specialPsidef}
\Psi(\tilde{r},t) = \Psi(w),\qquad w:= v\cdot \tilde{r} - \mu t + c,
\end{equation}
where $\mu$ and $c$ are real constants, and $v$ is a constant vector in $\mathbb{R}^{\ell m}$.
If we write 
\begin{equation}\label{vdef}
v:=\left[
v_1\,\,
v_2\,\,
\hdots\,\,
v_m
\right]^T,
\qquad
v_j:=
\left[
v_{1j},
v_{2j},
\hdots,
v_{\ell j}
\right]^T
\in\mathbb{R}^{\ell},\qquad 1\leq j\leq m,
\end{equation}
we find that
\begin{equation}\label{altwdef}
w := \sum_{s=1}^mv_s\cdot \tilde{r}_s- \mu t + c = \sum_{s=1}^m\sum_{i=1}^{\ell}v_{is}\tilde{x}_{is} -\mu t + c.
\end{equation}
We use the definition of $\Psi(w)$ provided by (\ref{altwdef}), along with the chain rule, 
to simplify the left side of (\ref{generalizedwaveequ}).
The chain rule implies that
\[
\frac{\partial\Psi}{\partial t} = \frac{d\Psi}{dw}\frac{\partial w}{\partial t} = -\mu\frac{d\Psi}{dw},
\]
which in turn implies that
\begin{equation}\label{specialcasePsipartial}
\frac{\partial^2\Psi}{\partial t^2}  = -\mu\frac{d^2\Psi}{dw^2}\frac{\partial w}{\partial t} = \mu^2\frac{d^2\Psi}{dw^2}.
\end{equation}
The chain rule also implies that
\[
\frac{\partial \Psi_{kj}}{\partial\tilde{x}_{is}} = \frac{d\Psi_{kj}}{dw}\frac{\partial w}{\partial \tilde{x}_{is}} 
= v_{is} \frac{d\Psi_{kj}}{dw},
\]
and by differentiating the above again with respect to $\tilde{x}_{is}$, we find that
\begin{equation}\label{specialcasetildepartial}
\frac{\partial^2 \Psi_{kj}}{\partial\tilde{x}_{is}^2} = v_{is}\frac{d^2\Psi_{kj}}{dw^2} \frac{\partial w}{\partial \tilde{x}_{is}} 
= v_{is}^2\frac{d^2\Psi_{kj}}{dw^2}.
\end{equation}
We substitute (\ref{specialcasetildepartial}) into (\ref{Laplacianop}) to obtain
\begin{equation}\label{SpecialLapop}
\Delta_{\tilde{r}}\Psi_{kj}(w) = \|v\|^2\frac{d^2\Psi_{kj}}{dw^2}.
\end{equation}
Equations (\ref{specialcasePsipartial}) and (\ref{SpecialLapop}) allow us to rewrite (\ref{generalizedwaveequ}) as
\begin{equation}\label{Specialcasgenwaveeq}
\mu^2\frac{d\Psi^2}{dw^2} - \|v\|^2\lambda^2\frac{d\Psi^2}{dw^2}
= \left[\mu^2 I_{pq\times pq} - \|v\|^2\lambda^2\right] \frac{d\Psi^2}{dw^2}= f(\Psi),
\end{equation}
where
$\lambda^2$ is the $pq\times pq$ matrix defined via (\ref{deltadef}) and $I_{pq\times pq}$ is the 
$pq\times pq$ identity matrix.
Observe that any solution $\Psi(w)$ to the nonlinear equation (\ref{Specialcasgenwaveeq}), with 
$w:= v\cdot \tilde{r} - \mu t + c$, is also a solution to the linear wave equation
\begin{equation}\label{secondwaveeq}
\|v\|^2\frac{\partial^2\Psi}{\partial t^2} - \mu^2 \Delta_{\tilde{r}}\Psi \equiv 0.
\end{equation}
The validity of (\ref{secondwaveeq}) follows from (\ref{specialcasePsipartial}) and (\ref{SpecialLapop}) 
and is the motivation behind our calling (\ref{Specialcasgenwaveeq}) 
the companion wave equation.
\begin{example}\label{secondexample}
Let $p = 3$, $q = 1$, $\ell = 3$, and $m$.  From (\ref{specialPsidef}) and (\ref{altwdef}) we see that
\[
\Psi(w):= [\Psi_1(w)],
\,\,\text{with}\,\,
\Psi_1(w) :=[\Psi_{11}(w),
\Psi_{21}(w),
\Psi_{31}(w)]^T,
\]
and
\begin{equation}\label{specficwdef1}
w := v\cdot\tilde{r}-\mu t + c = 
v_{11}\tilde{x}_{11} + v_{21}\tilde{x}_{21} + v_{31}\tilde{x}_{31} - \mu t + c,
\end{equation}
where we set
\[
\tilde{r} := [\tilde{r}_1],
\qquad\text{with}\,\,
\tilde{r}_1:= [\tilde{x}_{11}, \tilde{x}_{21},\tilde{x}_{31}]^T.
\]
This expression for $\Psi(w)$ means that (\ref{Specialcasgenwaveeq})  with 
\[
\lambda^2 := 
\left[
\begin{array}{ccc}
\lambda_1^2 & 0 & 0\\
0 & \lambda_1^2 & 0\\
0 & 0 &\lambda_1^2 
\end{array}
\right]
\]
is equivalent to
\begin{equation}\label{specialcase2vec}
\mu^2
\left[
\begin{array}{c}
\frac{d^2\Psi_{11}}{dw^2}
\vspace{0.3cm}\\
\frac{d^2\Psi_{21}}{dw^2}
\vspace{0.3cm}\\
\frac{d^2\Psi_{31}}{dw^2}
\end{array}\right]
-\|v\|^2
\left[
\begin{array}{c}
\lambda_1^2\frac{d^2\Psi_{11}}{dw^2}
\vspace{0.3cm}\\
\lambda_1^2\frac{d^2\Psi_{21}}{dw^2}\vspace{0.3cm}\\
\lambda_1^2\frac{d^2\Psi_{31}}{dw^2}\vspace{0.3cm}\\
\end{array}
\right]
 =
 \left[
\begin{array}{c}
f_{11}(\Psi_1,\Psi_2)
\vspace{0.3cm}\\
f_{21}(\Psi_1,\Psi_2)
\vspace{0.3cm}\\
f_{31}(\Psi_1,\Psi_2)
\end{array}
\right].
\end{equation}
\end{example}

\begin{example}\label{secondexampleA}
Let $p = 3$, $q = 2$, $\ell = 3$, and $m=1$.  From (\ref{specialPsidef}) and (\ref{altwdef}), we see that
\[
\Psi(w):= 
\left[
\begin{array}{c}
\Psi_1(w)\\
\Psi_2(w)
\end{array}
\right],
\,\,
\Psi_1(w) :=
\left[
\begin{array}{c}
\Psi_{11}(w)\\
\Psi_{21}(w)\\
\Psi_{31}(w)
\end{array}
\right],
\,\,
\Psi_2(w) :=
\left[
\begin{array}{c}
\Psi_{12}(w)\\
\Psi_{22}(w)\\
\Psi_{32}(w)
\end{array}
\right],
\]
and $w$ is defined as in Example \ref{secondexample}, namely
\[
w : = v\cdot\tilde{r}-\mu t + c = 
v_{11}\tilde{x}_{11} + v_{21}\tilde{x}_{21} + v_{31}\tilde{x}_{31}  - \mu t + c.
\]
The expression for $\Psi(w)$ means that (\ref{Specialcasgenwaveeq})  with 
\[
\lambda^2 := 
\left[
\begin{array}{cccccc}
\lambda_1^2 & 0 & 0 & 0 & 0 & 0\\
0 & \lambda_1^2 & 0  & 0 & 0 & 0\\
0 & 0 &\lambda_1^2 & 0 & 0 & 0\\
0 & 0 & 0 & \lambda_2^2 & 0 & 0\\
0 & 0 & 0 & 0 & \lambda_2^2 & 0\\
0 & 0 & 0 & 0 & 0 & \lambda_2^2
\end{array}
\right]
\]
is equivalent to
\begin{equation}\label{specialcase2vecA}
\mu^2
\left[
\begin{array}{c}
\frac{d^2\Psi_{11}}{dw^2}
\vspace{0.3cm}\\
\frac{d^2\Psi_{21}}{dw^2}
\vspace{0.3cm}\\
\frac{d^2\Psi_{31}}{dw^2}
\vspace{0.3cm}\\
\frac{d^2\Psi_{12}}{dw^2}\vspace{0.3cm}\\
\frac{d^2\Psi_{22}}{dw^2}\vspace{0.3cm}\\
\frac{d^2\Psi_{22}}{dw^2}
\end{array}\right]
-\|v\|^2
\left[
\begin{array}{c}
\lambda_1^2\frac{d^2\Psi_{11}}{dw^2}
\vspace{0.3cm}\\
\lambda_1^2\frac{d^2\Psi_{21}}{dw^2}\vspace{0.3cm}\\
\lambda_1^2\frac{d^2\Psi_{31}}{dw^2}\vspace{0.3cm}\\
\lambda_2^2\frac{d^2\Psi_{12}}{dw^2}\vspace{0.3cm}\\
\lambda_2^2\frac{d^2\Psi_{22}}{dw^2}\vspace{0.3cm}\\
\lambda_2^2\frac{d^2\Psi_{32}}{dw^2}
\end{array}
\right]
 =
 \left[
\begin{array}{c}
f_{11}(\Psi_1,\Psi_2)
\vspace{0.3cm}\\
f_{21}(\Psi_1,\Psi_2)
\vspace{0.3cm}\\
f_{31}(\Psi_1,\Psi_2)
\vspace{0.3cm}\\
f_{12}(\Psi_1,\Psi_2)\vspace{0.3cm}\\
f_{22}(\Psi_1,\Psi_2)\vspace{0.3cm}\\
f_{32}(\Psi_1,\Psi_2)
\end{array}
\right].
\end{equation}
\end{example}

\section{Traveling Waves for the Relative 2-Body and the 2-Body Problem}
In this section we look at special cases of Examples \ref{secondexample} and \ref{secondexampleA} associated with
the relative $2$-body and the $2$-body problem. 
 Let $r_i(t)\in\mathbb{R}^3$ be the position 
of the $i$th point mass $m_i$, where $i\in\{1,2\}$.  Newton's equations of celestial mechanics (NCME) for the $2$-body problem imply that
\begin{equation}\label{Newtonceleq}
r''_1(t) = \frac{Gm_2(r_2-r_1)}{\|r_2-r_1\|^3},\qquad r''_2(t) = \frac{Gm_1 (r_1-r_2)}{\|r_1-r_2\|^3},
\end{equation}
(recall that $\|v\|^2 = v^T\cdot v$).
By subtracting the two equations of (\ref{Newtonceleq}) we obtain the relative $2$-body problem.
\begin{equation}\label{2bodyrelsystem}
\frac{d^2\Delta_{12}}{dt^2}= -\frac{G(m_1+m_2)\Delta_{12}}{\|\Delta_{12}\|^3},\qquad \Delta_{12}(t):= r_1(t)-r_2(t).
\end{equation}
The companion wave equation of (\ref{2bodyrelsystem}) is a special case of Example \ref{secondexample}
 with
\begin{equation}\label{Psi1Delta12equiv}
\Psi_1(w)\equiv \Psi_{12}(w) = r_1(w)-r_2(w), \qquad \lambda_1^2\equiv \lambda_{12}^2 
\end{equation}
Then Equation (\ref{specialcase2vec}) becomes
\begin{equation}\label{specialcaseN12}
\mu^2
\left[
\begin{array}{c}
\frac{d^2\Psi_{1\,12}}{dw^2}
\vspace{0.3cm}\\
\frac{d^2\Psi_{2\,12}}{dw^2}
\vspace{0.3cm}\\
\frac{d^2\Psi_{3\,12}}{dw^2}
\end{array}
\right]
-\|v\|^2
\left[
\begin{array}{c}
\lambda_{12}^2\frac{d^2\Psi_{1\,12}}{dw^2}\vspace{0.3cm}\\
\lambda_{12}^2\frac{d^2\Psi_{2\,12}}{dw^2}\vspace{0.3cm}\\
\lambda_{12}^2\frac{d^2\Psi_{3\,12}}{dw^2}
\end{array}
\right]
=-G
\left[
\begin{array}{c}
\frac{(m_1+m_2)\Psi_{1\,12}(w)}{\|\Psi_{1\,12}(w)\|^3}
\vspace{0.3cm}\\
\frac{(m_1+m_2)\Psi_{2\,12}(w)}{\|\Psi_{2\,12}(w)\|^3}
\vspace{0.3cm}\\
\frac{(m_1+m_2)\Psi_{3\,12}(w)}{\|\Psi_{3\,12}(w)\|^3}
\end{array}
\right].
\end{equation}
By using the vector notation inherent in Example \ref{secondexample}, we efficiently rewrite as (\ref{specialcaseN12}) as
\begin{equation}\label{2bodywaverel}
[\mu^2 - \lambda_{12}^2\|v\|^2]\frac{d^2\Psi_{12}}{dw^2} = -\frac{G(m_1+m_2)\Psi_{12}}{\|\Psi_{12}\|^3}.
\end{equation}
Our goal is to find a closed form solution of (\ref{2bodywaverel}).  To that end assume $\Psi_{12}(w) = w^{\frac{2}{3}}S_1$ 
for some nonzero constant vector in $\mathbb{R}^3$.  If we substitute this form of 
$\Delta_{12}(w)$ into (\ref{2bodywaverel}) we discover that
\[
-\frac{2}{9}[\mu^2 - \lambda_{12}^2\|v\|^2]S_1 = -\frac{G(m_1+m_2)}{\|S_1\|^3}S_1.
\]
By taking the norm of both sides of preceding equation we prove the following proposition.
\begin{proposition}\label{2bodyrelprop}
Given the 2-body relative difference companion wave equation
\[
[\mu^2 - \lambda_{12}^2\|v\|^2]\frac{d^2\Psi_{12}}{dw^2} = -\frac{G(m_1+m_2)\Psi_{12}}{\|\Psi_{12}\|^3},
\qquad \mu^2 - \lambda_{12}^2\|v\|^2\neq 0,
\]
where $\mu,\lambda\in\mathbb{R}/\{0\}$ and $v\in\mathbb{R}^3/\{\overrightarrow{0}\}$,
choose any $S_1\in\mathbb{R}^3/\overrightarrow{0}$ whose norm satisfies
\begin{equation}\label{2bodywaverela}
\|S_1\|^3 = \frac{9G(m_1+m_2)}{2|\mu^2 - \lambda_{12}^2\|v\|^2|}.
\end{equation}
Then $\Psi_{12}(w) = w^{\frac{2}{3}}S_1$ is a solution to (\ref{2bodywaverela}) 
whenever $w\in(a,b)$, where $0\not\in(a,b)$.
\end{proposition}

\begin{remark}
Proposition \ref{2bodyrelprop} is valid if $\|v\| = 0$, in which case (\ref{2bodywaverela}) reduces to
\[
\|S_1\| = \left(\frac{9G(m_1+m_2)}{2\mu^2}\right)^{\frac{1}{3}}.
\]
If in addition $\mu = -1$, then $w = t$ and (\ref{2bodywaverel}) is the Newtonian relative difference system (\ref{2bodyrelsystem})
with $\Delta_{12}(t)\equiv r_1(t)-r_2(t)$.  In this case, Proposition \ref{2bodyrelprop} implies that for any unit vector $U$,
\[ 
\Delta_{12}(t) = t^{\frac{2}{3}}\|S_1\|U,\qquad \|S_1\| = \left(\frac{9G(m_1+m_2)}{2}\right)^{\frac{1}{3}}
\]
is a solution to (\ref{2bodyrelsystem}) whenever $t\in(a,b)$, where $0\not\in(a,b)$.
\end{remark}

\begin{remark}
We think about the vector $\left<w^{\frac{2}{3}}, \|S_1\|U\right>$ as a salient feature of the wave.
The component  $w^{\frac{2}{3}}$ provides information on the wave front as seen by the discussion in the
next section. The scalar $\|S_1\|\neq 0$ is uniquely determined and could be interpreted as a measure of the
amplitude. In our interpretation of Proposition \ref{2bodyrelprop} we look at a union of solutions
$\Psi_{12}(w) =w^{\frac{2}{3}}\|S_1\|U$, where $U$ is any unit vector in $\mathbb{R}^3$.
\end{remark}

Next we look at a special case of Example \ref{secondexampleA}, namely the company wave equation to
(\ref{Newtonceleq}). If we set 
\begin{equation}\label{psieqivr}
\Psi_1(w)\equiv r_1(w),\qquad \Psi_2(w)\equiv r_2(w),
\end{equation}
Equation  (\ref{specialcase2vecA}) becomes
\begin{equation}\label{specialcaseN2}
\mu^2
\left[
\begin{array}{c}
\frac{d^2\Psi_{11}}{dw^2}
\vspace{0.3cm}\\
\frac{d^2\Psi_{21}}{dw^2}
\vspace{0.3cm}\\
\frac{d^2\Psi_{31}}{dw^2}
\vspace{0.3cm}\\
\frac{d^2\Psi_{12}}{dw^2}\vspace{0.3cm}\\
\frac{d^2\Psi_{22}}{dw^2}\vspace{0.3cm}\\
\frac{d^2\Psi_{22}}{dw^2}
\end{array}\right]
-\|v\|^2
\left[
\begin{array}{c}
\lambda_1^2\frac{d^2\Psi_{11}}{dw^2}
\vspace{0.3cm}\\
\lambda_1^2\frac{d^2\Psi_{21}}{dw^2}\vspace{0.3cm}\\
\lambda_1^2\frac{d^2\Psi_{31}}{dw^2}\vspace{0.3cm}\\
\lambda_2^2\frac{d^2\Psi_{12}}{dw^2}\vspace{0.3cm}\\
\lambda_2^2\frac{d^2\Psi_{22}}{dw^2}\vspace{0.3cm}\\
\lambda_2^2\frac{d^2\Psi_{32}}{dw^2}
\end{array}
\right]
=G
\left[
\begin{array}{c}
\frac{m_2(\Psi_{12}(w)-\Psi_{11}(w))}{\|\Psi_{2}(w)-\Psi_{1}(w)\|^3}
\vspace{0.3cm}\\
\frac{m_2(\Psi_{22}(w)-\Psi_{21}(w))}{\|\Psi_{2}(w)-\Psi_{1}(w)\|^3}
\vspace{0.3cm}\\
\frac{m_2(\Psi_{32}(w)-\Psi_{31}(w))}{\|\Psi_{2}(w)-\Psi_{11}(w)\|^3}
\vspace{0.3cm}\\
\frac{m_1(\Psi_{11}(w)-\Psi_{12}(w))}{\|\Psi_1(w)-\Psi_2(w)\|^3}
\vspace{0.3cm}\\
\frac{m_1(\Psi_{21}(w)-\Psi_{22}(w))}{\|\Psi_1(w)-\Psi_2(w)\|^3}
\vspace{0.3cm}\\
\frac{m_1(\Psi_{31}(w)-\Psi_{32}(w))}{\|\Psi_1(w)-\Psi_2(w)\|^3}
\vspace{0.3cm}\\
\end{array}
\right].
\end{equation}
Using the notation of Example \ref{secondexample} we efficiently rewrite (\ref{specialcaseN2})
 as the following two $3\times 1$ nonlinear vector differential equations:
\begin{align}\label{specialcaseN2vecform}
\left[\mu^2 - \lambda_1^2\|v\|^2\right]\frac{d^2\Psi_1}{dw^2} &= \frac{Gm_2(\Psi_2- \Psi_1)}{\|\Psi_2-\Psi_1\|^3}\notag\\
\left[\mu^2 - \lambda_2^2\|v\|^2\right]\frac{d^2\Psi_2}{dw^2} &= \frac{Gm_1(\Psi_1- \Psi_2)}{\|\Psi_2-\Psi_1\|^3}.
\end{align}
We will derive closed form solutions to the preceding system.  To that end
we assume there exists two {\it distinct} nonzero constant $3\times 1$ vectors $S_1$ and $S_2$ such that
\begin{equation}\label{explicitsol}
\Psi_1(w) = w^{\frac{2}{3}}S_1,\qquad \Psi_2(w) = w^{\frac{2}{3}}S_2.
\end{equation}
If we place the expressions of (\ref{explicitsol}) into (\ref{specialcaseN2vecform}) we find that
\begin{align}\label{specialcaseN2explictform}
-\frac{2}{9}\left[\mu^2 - \lambda_1^2\|v\|^2\right]S_1 &=  \frac{Gm_2(S_2- S_1)}{\|S_2-S_1\|^3}\notag\\
-\frac{2}{9}\left[\mu^2 - \lambda_2^2\|v\|^2\right]S_2 & =\frac{Gm_1(S_1- S_2)}{\|S_2-S_1\|^3}.
\end{align}
We may rewrite (\ref{specialcaseN2explictform}) as
\begin{align}\label{S1S2solve}
-\frac{2}{9}S_1 &= \theta_1\frac{S_2- S_1}{\|S_2-S_1\|^3},\qquad \theta_1:= \frac{Gm_2}{\mu^2 - \lambda_1^2\|v\|^2}\neq 0,\notag\\
-\frac{2}{9}S_2 &= \theta_2\frac{S_1- S_2}{\|S_2-S_1\|^3},\qquad \theta_2= \frac{Gm_1}{\mu^2 - \lambda_2^2\|v\|^2}\neq 0.
\end{align}
Because $S_1\neq S_2$, $\|S_1-S_2\|\neq 0$ and the equations of (\ref{S1S2solve}) are well defined.
If we subtract the second equation of (\ref{S1S2solve})  from the first we discover that
\begin{equation}\label{firstsub}
-\frac{2}{9}(S_1-S_2) = (\theta_1+\theta_2)\frac{S_2- S_1}{\|S_2-S_1\|^3}.
\end{equation}
Without loss of generality, we can assume $\theta_1 + \theta_2\neq 0$.  
For if $\theta_1 + \theta_2 = 0$, then $\theta_1 = -\theta_2$ and
(\ref{S1S2solve}) becomes
\[
-\frac{2}{9}S_1 = \theta_2\frac{S_1- S_2}{\|S_2-S_1\|^3},\qquad 
-\frac{2}{9}S_2 = \theta_2\frac{S_1- S_2}{\|S_2-S_1\|^3},
\]
which implies that $S_1 = S_2$, a contradiction to the assumption that $S_1\neq S_2$.
Note that $\theta_1 + \theta_2\neq 0$ implies that
\begin{equation}\label{restriction1}
-\frac{m_2}{m_1}\neq \frac{\mu^2 - \lambda_1^2\|v\|^2}{\mu^2 - \lambda_2^2\|v\|^2}.
\end{equation}
\begin{remark}
Mathematically, $\theta_1 + \theta_2\neq 0$ is a bona fide restriction.  However, 
we have the flexibility to choose the parameters $\lambda_1$ and $\lambda_2$ in
such a way as to ensure the validity of Inequality (\ref{restriction1}), and 
hence we can avoid this restriction.  Furthermore, if $\theta_1$ and $\theta_2$ have the same sign,
then automatically $\theta_1+\theta_2\neq 0$.
\end{remark}
If we take the norm of both sides of (\ref{firstsub}) we find that
\begin{equation}\label{normresult}
\frac{1}{\|S_1-S_2\|^3} = \frac{2}{9|\theta_1+\theta_2|}.
\end{equation}
We then substitute (\ref{normresult}) into (\ref{S1S2solve}) to obtain
\begin{align}\label{simplifiedS1S2solve}
-\frac{2}{9}S_1 &= \frac{2\theta_1}{9|\theta_1+\theta_2|}(S_2-S_1)\notag\\
-\frac{2}{9}S_2 &= \frac{2\theta_2}{9|\theta_1+\theta_2|}(S_1-S_2),
\end{align}
which is equivalent to the linear system
\begin{align}\label{linearsys}
\left(1-\frac{\theta_1}{|\theta_1 + \theta_2|}\right)S_1 + \frac{\theta_1}{|\theta_1+\theta_2|}S_2 &= \overrightarrow{0}\notag\\
-\frac{\theta_2}{|\theta_1+\theta_2|}S_1 + \left(\frac{\theta_2}{|\theta_1+\theta_2|}-1\right)S_2&= \overrightarrow{0}.
\end{align}
If $\theta_1 + \theta_2 > 0$, then $|\theta_1 + \theta_2| = \theta_1 + \theta_2$, and we may
 simplify the coefficients in the round brackets of (\ref{linearsys}) to obtain
\begin{align}\label{lienarsyssimp}
\theta_2S_1  +  \theta_1S_2 &= \overrightarrow{0}\notag\\
-\theta_2S_1 - \theta_1S_2 &= \overrightarrow{0}.
\end{align}
System (\ref{lienarsyssimp}) has only one independent equation from which we deduce that
\begin{equation}\label{finalreduction}
S_1 = -\frac{\theta_1}{\theta_2}S_2.
\end{equation}
Now assume that $\theta_1 + \theta_2 < 0$.  Then $|\theta_1 + \theta_2| = -(\theta_1+\theta_2)$ and (\ref{linearsys}) becomes
\begin{align}\label{linearsysneg}
\left(1+\frac{\theta_1}{\theta_1 + \theta_2}\right)S_1 - \frac{\theta_1}{\theta_1+\theta_2}S_2 &= \overrightarrow{0}\notag\\
\frac{\theta_2}{\theta_1+\theta_2}S_1 - \left(\frac{\theta_2}{\theta_1+\theta_2}+1\right)S_2&= \overrightarrow{0}.
\end{align}
System (\ref{linearsysneg}) is equivalent to 
\begin{align}\label{linearsysnegsimp}
(2\theta_1+\theta_2)S_1 - \theta_1S_2 &= \overrightarrow{0}\notag\\
\theta_2S_1 -(\theta_1 + 2\theta_2)S_2 &= \overrightarrow{0},
\end{align}
If we subtract the second equation of (\ref{linearsysnegsimp}) from the first we obtain
\[
2\theta_1S_1 + 2\theta_2S_2 = \overrightarrow{0},
\]
which implies that
\begin{equation}\label{secondfinalsol}
S_1 = -\frac{\theta_2}{\theta_1}S_2.
\end{equation}
We then substitute (\ref{secondfinalsol}) into the first equation of (\ref{linearsysnegsimp}) to obtain
\begin{equation}\label{coeffeq1}
\frac{(2\theta_1+\theta_2)\theta_2}{\theta_1} + \theta_1 = 0,
\end{equation}
since we are implicitly assuming that $S_2\neq \overrightarrow{0}$.
However, by multiplying both sides of (\ref{coeffeq1}) by $\theta_1\neq 0$, we obtain
\[
(2\theta_1+\theta_2)\theta_2 + \theta_1^2 = (\theta_1+\theta_2)^2 = 0,
\]
and this implies that $\theta_1+ \theta_2 = 0$, a contradiction to our 
requirement that $S_1\neq S_2$.

\medskip
In summary we have proven the following proposition.
\begin{proposition}\label{Newton2bodyprop}
Let $\mu,\lambda_{1}, \lambda_{2}\in \mathbb{R}/\{0\}$ be arbitrarily chosen.
Choose any nonzero vector $v\in\mathbb{R}^3$ such that
\[
\mu^2 -\lambda_1^2\|v\|^2\neq 0,\qquad \mu^2 -\lambda_2^2\|v\|^2\neq 0.
\]
Define
\[
\theta_1:=\frac{Gm_2}{\mu^2 -\lambda_1^2\|v\|^2}\neq 0 ,\qquad \theta_2:= \frac{Gm_1}{\mu^2 -\lambda_2^2\|v\|^2}\neq 0.
\]
We require that $\theta_1 + \theta_2 >0$.
This is equivalent to requiring that $\lambda_1$ and $\lambda_2$ are chosen to satisfy
\[
\frac{m_2}{m_1} > -\frac{\mu^2-\lambda_1^2\|v\|^2}{\mu^2 - \lambda_2^2\|v\|^2}.
\]
Define $S_1 := -\theta_1/\theta_2S_2$, where $S_2$ is a 
nonzero vector in $\mathbb{R}^3$ whose norm satisfies
\begin{equation}\label{normdiffreq}
\frac{1}{\|S_1-S_2\|^3} = \frac{2}{9|\theta_1+\theta_2|}.
\end{equation}
 Then $\Psi_1(w) = -\theta_1/\theta_2w^{\frac{2}{3}}S_2$ and $\Psi_2(w) = w^{\frac{2}{3}}S_2$ are solutions to 
\begin{align}\label{propositionsys}
\left[\mu^2 - \lambda_1^2\|v\|^2\right]\frac{d^2\Psi_1}{dw^2} &= \frac{Gm_2(\Psi_2- \Psi_1)}{\|\Psi_2-\Psi_1\|^3}\notag\\
\left[\mu^2 - \lambda_2^2\|v\|^2\right]\frac{d^2\Psi_2}{dw^2} &= \frac{Gm_1(\Psi_1- \Psi_2)}{\|\Psi_2-\Psi_1\|^3},
\end{align}
whenever $w\in(a,b)$, where $0\not\in(a,b)$.
\end{proposition}

\begin{remark}  
A careful reading of the proof of Proposition \ref{Newton2bodyprop} shows that
Proposition \ref{Newton2bodyprop} is valid if $v = \overrightarrow{0}$, in which case
\[
\theta_1 = G\mu^{-2}m_2 > 0,\qquad \theta_2 = G\mu^{-2}m_1 > 0.
\]
\end{remark}

If $\|v\|= 0$, $\mu = -1$, and $c = 0$, we find that $w = t$.  Then (\ref{propositionsys}) reduces to 
NCME, namely (\ref{Newtonceleq}).
In this case $\theta_1 := Gm_2$, $\theta_2 := Gm_1$, and Proposition \ref{Newton2bodyprop} implies that for any unit vector
$U$,
\begin{equation}\label{newtonclosedformsol}
\Psi_1(t) \equiv r_1(t) =  -m_2/m_1t^{\frac{2}{3}}\|S_2\|U,\qquad \Psi_2(t)\equiv r_2(t) = t^{\frac{2}{3}}\|S_2\|U,
\quad \text{$t\in(a,b)$, $0\not\in(a,b)$},
\end{equation}
are solution to (\ref{Newtonceleq}).  Note that $\|S_2\|$ is determined via (\ref{normdiffreq}).  Since 
$S_1 := -\theta_1/\theta_2S_2 = -m_2/m_1S_2$, Equation (\ref{normdiffreq}) becomes
\[
\frac{1}{\left[\frac{m_2}{m_1} + 1\right]^3\|S_2\|^3} = \frac{2}{9G(m_1+m_2)},
\]
which implies that
\begin{equation}\label{vzeroS2norm}
\|S_2\|^3 = \frac{9G(m_1+m_2)}{2\left[\frac{m_2}{m_1} + 1\right]^3} = \frac{9Gm_1^3}{2(m_1+m_2)^2}.
\end{equation}
If we substitute (\ref{vzeroS2norm}) into (\ref{newtonclosedformsol}) we find that
\begin{equation}\label{newtonclosedformsolA}
r_1(t)  = -\frac{\hat{\gamma}m_2}{(m_1+m_2)^{\frac{2}{3}}}t^{\frac{2}{3}}U,\qquad 
r_2(t) = -\frac{\hat{\gamma}m_1}{(m_1+m_2)^{\frac{2}{3}}}t^{\frac{2}{3}}U,
\quad 
\hat{\gamma}:=\left(\frac{9G}{2}\right)^{\frac{1}{3}}.
\end{equation}
As $t\rightarrow 0$, $r_1(t) = r_2(t) = 0$ and the two point masses collide.  At the time of collision, 
the point masses emit a gravitational wave field which is modeled by the companion wave equation 
(\ref{propositionsys}), where $w = v\cdot \tilde{r} - \mu t + c$ with $\|v\|\neq 0$.
If the gravitational wave field travels through homogeneous space, we set $\lambda_{1}^2 = \lambda_2^2$
and $v = \pm 1/\sqrt(3)(1,1,1)^T$,
in which case
\[
\theta_1:=\frac{Gm_2}{\mu^2 -\lambda_1^2\|v\|^2},\qquad \theta_2:=\frac{Gm_1}{\mu^2 -\lambda_1^2\|v\|^2}.
\]
Then $-\theta_1/\theta_2 = -m_2/m_1$ as in the case NCME.
Hence (\ref{vzeroS2norm}) is valid and we deduce that (\ref{newtonclosedformsolA}), with $t$ replaced by $w$, $r_1(t)$ replaced with $\Psi_1(w)$, 
and $r_2(t)$ replaced by $\Psi_2(w)$, are the
closed form solutions for (\ref{propositionsys}).

\section{The Fronts of the Relative 2-Body Problem}
\begin{figure}[H]
\begin{center}
\includegraphics[height=1.9in,width=4.8in]{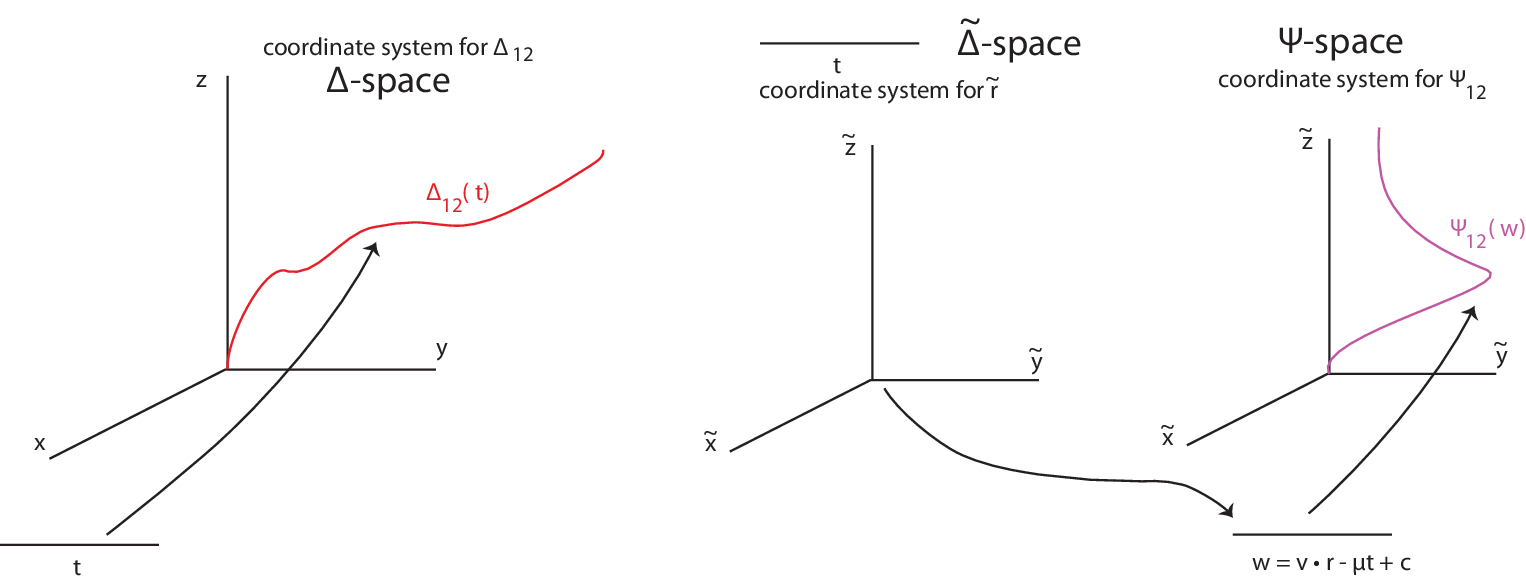}
\end{center}
\caption{Three copies of $\mathbb{R}^3$ for the relative $2$-body problem and its
companion wave equation.}
\label{coordinateaxis}
\end{figure}

\begin{figure}[H]
\begin{center}
\includegraphics[height=2.3in,width=1.8in]{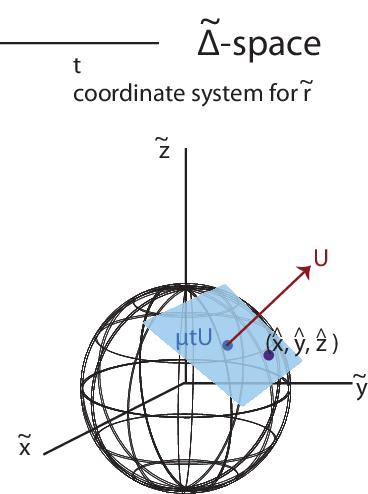}
\end{center}
\caption{The tangent plane to the sphere $\tilde{x}^2 + \tilde{y}^2 + \tilde{z}^2 = (\mu t)^2$ at the point $\mu t U$.}
\label{tangentplanesphere}
\end{figure}

\begin{figure}[H]
\begin{center}
\includegraphics[height=1.8in,width=2.3in]{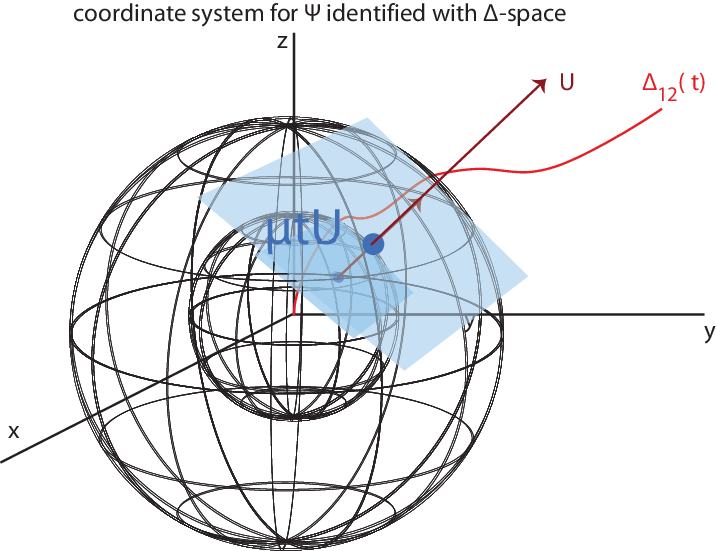}
\end{center}
\caption{Gravitational wave front for a collision at $(0,0,0)$ as tangent planes
to growing spheres.}
\label{movetangentplanes}
\end{figure}

In this last section we provide a physical interpretation for Proposition \ref{2bodyrelprop} which describes the shape 
of the gravitational wave fronts associated
with the collision of a binary star. For this interpretation we need to consider three independent copies of $\mathbb{R}^3$.
The first copy of $\mathbb{R}^3$ is associated with $\Delta_{12}(t):=r_1(t)-r_2(t)$, where $\Delta_{12}(t)$ is a solution to
the relative difference equation implied by NCME, namely
\begin{equation}\label{rewriteNewrel2body}
\frac{d^2\Delta_{12}}{dt^2} = -\frac{G(m_1+m_2)\Delta_{12}}{\|\Delta_{12}\|^3}.
\end{equation}
The coordinates for $\Delta_{12}(t)$ are the standard Euclidean coordinates denoted  by an $x$-axis, a $y$-axis, and a $z$-axis.
We will call this copy of $\mathbb{R}^3$ the $\Delta$-{\it space}.
The second copy of $\mathbb{R}^3$ is associated with $\Psi_{12}(w):=w^{2/3}S_1$, where 
according to Proposition \ref{2bodyrelprop},
$\Psi_{12}(w)$ is a closed form solution to the companion wave equation
\begin{equation}\label{rewritecompwaveeq}
[\mu^2 - \lambda_{12}^2\|v\|^2]\frac{d^2\Psi_{12}}{dw^2} = -\frac{G(m_1+m_2)\Psi_{12}}{\|\Psi_{12}\|^3}.
\end{equation}
The coordinates for $\Psi_{12}(w)$ are also the standard Euclidean coordinates denoted by an $\tilde{x}\equiv\tilde{x}_{11}$-axis, 
a $\tilde{y}\equiv\tilde{x}_{21}$-axis, 
and a $\tilde{z}\equiv\tilde{x}_{31}$-axis.  This copy of $\mathbb{R}^3$ is referred to as the $\Psi$-{\it space}.
Recall from (\ref{specficwdef1}), (with $v_{11}\equiv v_1$, $v_{21}\equiv v_2$, and $v_{31}\equiv v_3$),
that $w = v_{1}\tilde{x} + v_{2}\tilde{y} + v_{3}\tilde{z}-\mu t + c = \tilde{r}\cdot v - \mu t+ c$.
The third copy of $\mathbb{R}^3$ is for $\tilde{r}$ and is associated with the domain of
$\Psi_{12}:\mathbb{R}^4\rightarrow R^3$ when $\Psi_{12}(w)$ is interpreted as the composite function
$\Psi_{12}(\tilde{r},t)$. The coordinates of $\tilde{r}$ are also represented by the $\tilde{x}$-axis, the $\tilde{y}$-axis, 
and the $\tilde{z}$-axis; see Figure \ref{coordinateaxis}. 
We will refer to the copy of $\mathbb{R}^3$ associated with $\tilde{r}$, when supplemented by $t$,
as the $\tilde{\Delta}$-{\it space}.  
Observe that the identity map is a continuously differentiable bijection between the copy of $\mathbb{R}^3$ in
the $\tilde{\Delta}-space$
and the $\Psi$-space.
We assume that at $t = 0$ and when $(\tilde{x},\tilde{y},\tilde{z}) = (0,0,0)$ in the $\Psi$-space, there is a collision between the
two point masses represented by $r_1(t)$ and $r_2(t)$, namely that $\Delta_{12}(0) = (0,0,0)$. This collision creates a gravitational wave front in
the $\Psi$-space.
 We want to describe the shape and position (as a function of time) of this gravitational wave in the $\Delta$-space.  In order to answer this question
 we make the following definition
 \begin{definition}\label{frontlocdef}
 The front in the $\Delta$-space occurs when the gradient
 $\|\left(\nabla_{(\tilde{x},\tilde{y},\tilde{z})}\Psi_{j\,12}(w),\partial \Psi_{j\,12}/\partial t\right)\|$ 
 is undefined and/or unbounded for some $j$ where $1\leq j\leq 3$.
 \end{definition}
 Since $\Psi_{12}(w):=w^{2/3}S_1$, where $w = v_1\tilde{x} + v_2\tilde{y} + v_3\tilde{z}-\mu t + c$, and since
 $S_1:= [s_{11}, s_{21},s_{31}]^T$, we find that
 \begin{equation}\label{gradientvect}
\left(\nabla_{(\tilde{x},\tilde{y},\tilde{z})}\Psi_{j\,12}(w),\frac{\partial \Psi_{j\,12}}{\partial t}\right)
=
\left(\frac{2}{3}w^{-\frac{1}{3}}(v_1,v_2,v_3), -\frac{2}{3}w^{-\frac{1}{3}}\mu\right)s_{j1}.
\end{equation}
Because the conditions of Proposition \ref{2bodyrelprop} require that $v\neq \overrightarrow{0}$ and $\mu\neq 0$, and since 
a-priori $s_{j1}neq 0$ for all $1\leq j\leq 3$, the calculation of
(\ref{gradientvect}) implies that the gradient of Definition \ref{frontlocdef} is unbounded if $w =0$, which is equivalent to
requiring that $v_1\tilde{x} + v_2\tilde{y} + v_3\tilde{z} = \mu t -c$ in the $\tilde{\Delta}$-space.  We have the flexibility to choose $v = (v_1,v_2,v_3)$, 
to choose $c$, and according to
 Proposition \ref{2bodyrelprop}, we also have the freedom to choose the direction of $S_1$, namely $U = S_1/\|S_1\|$. So we set $c = 0$ and
 $v = U = (u_1,u_2, u_3)$.  Then the gradient of Definition \ref{frontlocdef} is unbounded if $u_1\tilde{x} + u_2\tilde{y} + u_3\tilde{z} = \mu t$.
 For fixed $\mu  >0$ and fixed $t> 0$, by varying the unit vector $U\in\mathbb{R}^3$, 
 the equations $u_1\tilde{x} + u_2\tilde{y} + u_3\tilde{z} = \mu t$  in the $\tilde{\Delta}$-space
form a collection of planes which are tangent to the sphere $S^2_{\mu t}$, where
\[
S^2_{\mu t}:= \{(\tilde{x},\tilde{y},\tilde{z})\mid \tilde{x}^2 + \tilde{y}^2 + \tilde{z}^2 = (\mu t)^2\}.
\]
To prove this last statement consider the point $\mu tU = \mu t(u_1, u_2,u_3) \in S^2_{\mu t}$.  
The tangent plane at $\mu tU$  is given by
the equation
\[
U\cdot (\hat{x}-\mu t u_1, \hat{y}-\mu t u_2,\hat{z}-\mu t u_3) =0,
\]
where $(\hat{x},\hat{y},\hat{z})$ is an arbitrary point on the tangent plane.
The above equation is equivalent to 
\[
u_1\hat{x} + u_2\hat{y} + u_3\hat{z} =  \mu t(u_1^2 + u_2^2 + u_3^2) = \mu t\|U\|^2 = \mu t,
\]
which after an appropriate renaming of the variables becomes $u_1\tilde{x} + u_2\tilde{y} + u_3\tilde{z} = \mu t$;
see Figure \ref{tangentplanesphere}.
The final step is to use the identity map between the copy of $\mathbb{R}^3$ from the $\tilde{\Delta}$-space 
and the $\Psi$-space and  identify the $\Delta$-space with the $\Psi$-space.
 Observe that for fixed $\mu$ and fixed $t>0$, 
 the planes tangent to the sphere $x^2 + y^2 + z^2 = (\mu t)^2$  are 
 representations of the gravitational wave front at time $t$ which results from the collision $\Delta_{12}(0) = (0,0,0)$.
 As $t$ increases, the sphere expands to fill in the $\Delta$-space. For fixed $\mu$, fixed positive $t$, and fixed $U$, the plane
 $u_1x + u_2y + u_3z = \mu t$  is tangent to the sphere at the point $\mu t U = \mu t(u_1, u_2, u_3)$.  The normal line through the point  $\mu t U$ has
 equation 
 \[
 x(s) = t\mu u_1 + su_1,\qquad y(s) = t\mu u_2 + su_2, \qquad z(s) = t\mu u_2 + su_3.
 \]
 By increasing $s$, we move the tangent plane along this normal line and obtain the plane  $u_1x + u_2y + u_3z = \mu(t+s)$ which is
 tangent to the sphere with equation $x^2 + y^2 + z^2 = (\mu (t+s))^2$; see Figure \ref{movetangentplanes}.

\begin{figure}[ht]
\begin{center}
\includegraphics[height=1.9in,width=4.8in]{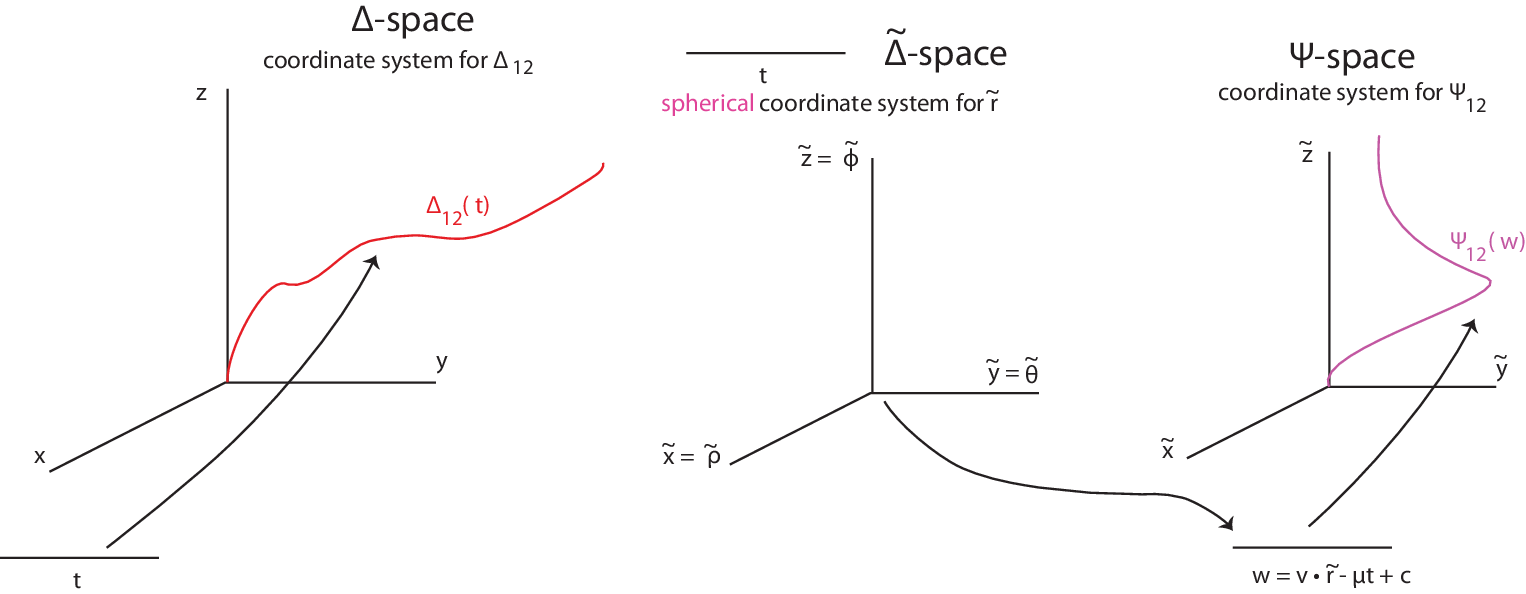}
\end{center}
\caption{Spherical coordinates for the $\tilde{\Delta}$-space.}
\label{sphericalframespace}
\end{figure}

\begin{figure}[ht]
\begin{center}
\includegraphics[height=3.2in,width=5in]{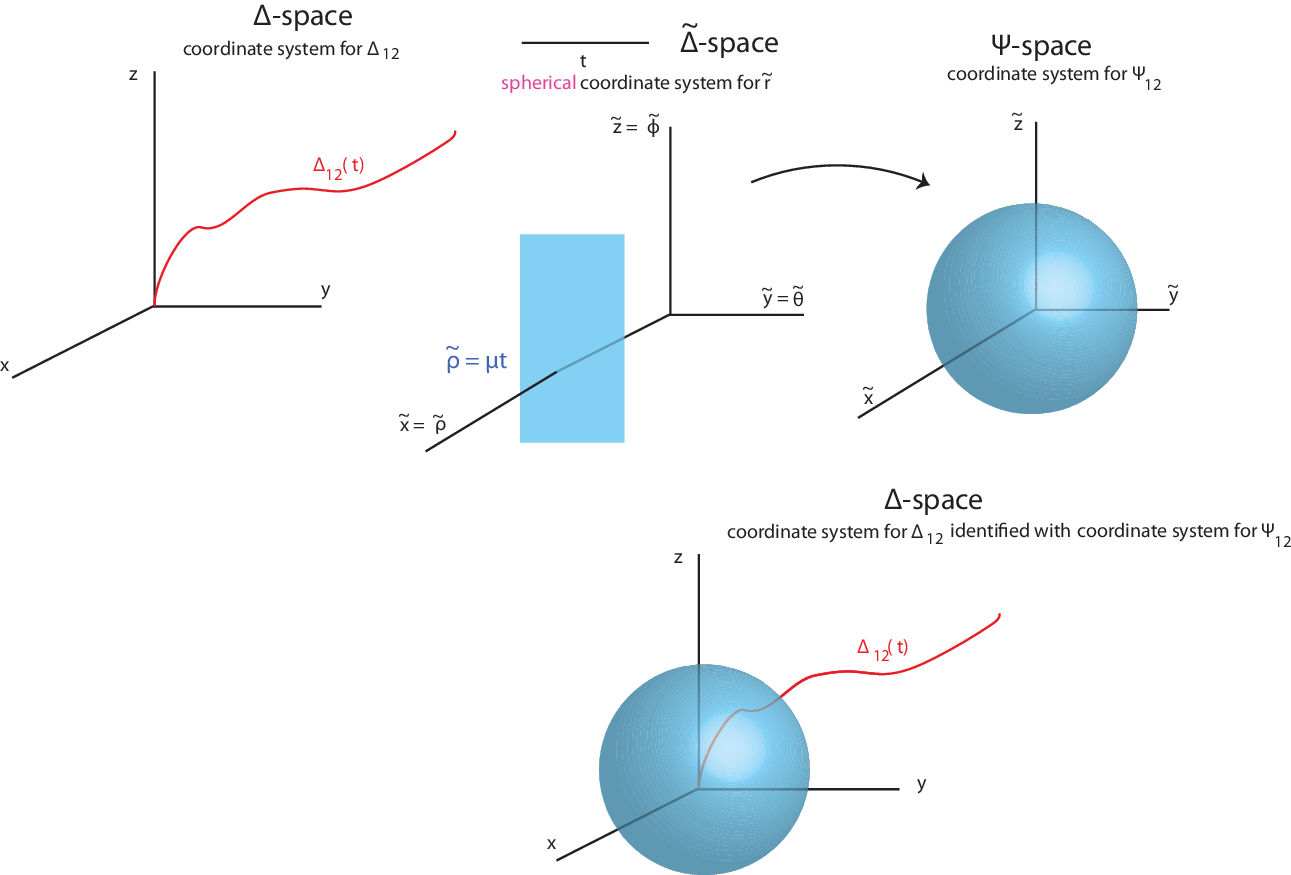}
\end{center}
\caption{Gravitational wave front in the direction $(1,0,0)$ resulting from spherical coordinates.}
\label{sphericalwavefront}
\end{figure}

\medskip
So far we have required that the $\Delta$-space, the $\Psi$-space, and the $\tilde{\Delta}$-space are represented via Cartesian coordinates.
We obtain other gravitational wave fronts if we allow more general coordinate systems for the $\tilde{\Delta}$-space.
For example, we could represent the $\tilde{\Delta}$-space in spherical coordinates $(\tilde{\rho}, \tilde{\theta},\tilde{\varphi})$.
See Figure \ref{sphericalframespace}. The coordinates for $\Psi_{12}(w)$ are still the Euclidean coordinates given by
$(\tilde{x},\tilde{y}, \tilde{z})$.
Once again when $(\tilde{x},\tilde{y},\tilde{z}) = (0,0,0)$ in the $\Psi$-space and when $t = 0$, there is a collision
between the two point masses in the $\Delta$-space, i.e. $\Delta_{12}(0) = (0,0,0)$. Definition
\ref{frontlocdef}  and Proposition \ref{2bodyrelprop} are still valid.  For the case of spherical coordinates, 
since $\tilde{x} = \tilde{\rho}$, $\tilde{y} = \tilde{\theta}$ and $\tilde{z} =\tilde{\varphi}$, Equation (\ref{gradientvect})
is still valid and the gradient of
Definition \ref{frontlocdef} is unbounded if $w =0$.
However, since $w = v\cdot\tilde{r} -\mu t-c = v_1\tilde{\rho} + v_2\tilde{\theta} + v_3\tilde{\varphi} -\mu t +c$, 
we find that $w = 0$ if and only if $v_1\tilde{\rho} + v_2\tilde{\theta} + v_3\tilde{\varphi} = \mu t - c$.  As before,
we have the flexibility to choose $v$, $c$ and $U = S_1/\|S_1\|$. Thus we set $c = 0$, and $u = v = (1,0,0)$.
Then $v_1\tilde{\rho} + v_2\tilde{\theta} + v_3\tilde{\varphi} = \mu t - c$ simplifies to $\tilde{\rho} = \mu t$.
For fixed $\mu$ and fixed $t > 0$, the image of $\tilde{\rho}$ in the $\Psi$-space is the sphere 
$S^2_{\mu t}$.
We then identify the $\Delta$-space with the $\Psi$-space and note that
the gravitational wave front at time $t$ in the direction $(1,0,0)$ is $S^2_{\mu t}$.  As $t$ increases the radius of the sphere 
increases to fill the $\Delta$-space; see Figure \ref{sphericalwavefront}.

\begin{figure}[H]
\begin{center}
\includegraphics[height=1.9in,width=4.8in]{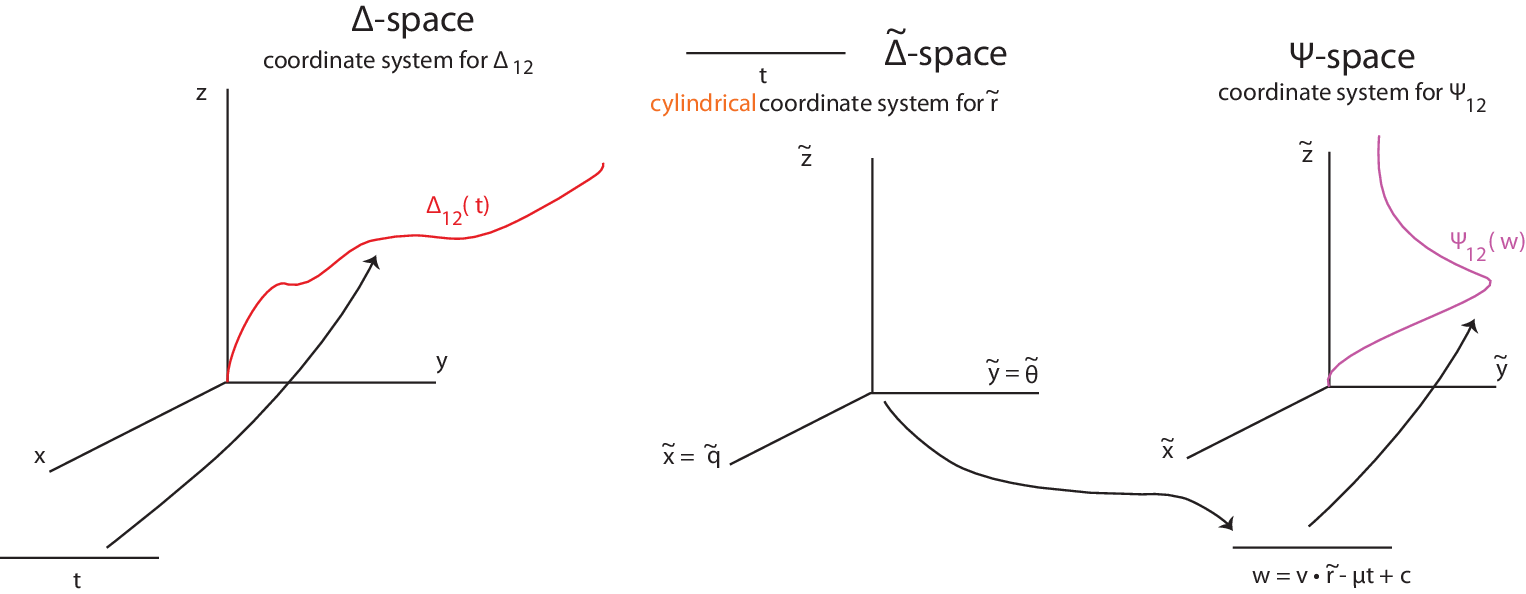}
\end{center}
\caption{Cylindrical coordinates for the $\tilde{\Delta}$-space.}
\label{cyclindricalcoordsys}
\end{figure}

\begin{figure}[H]
\begin{center}
\includegraphics[height=2.8in,width=4.8in]{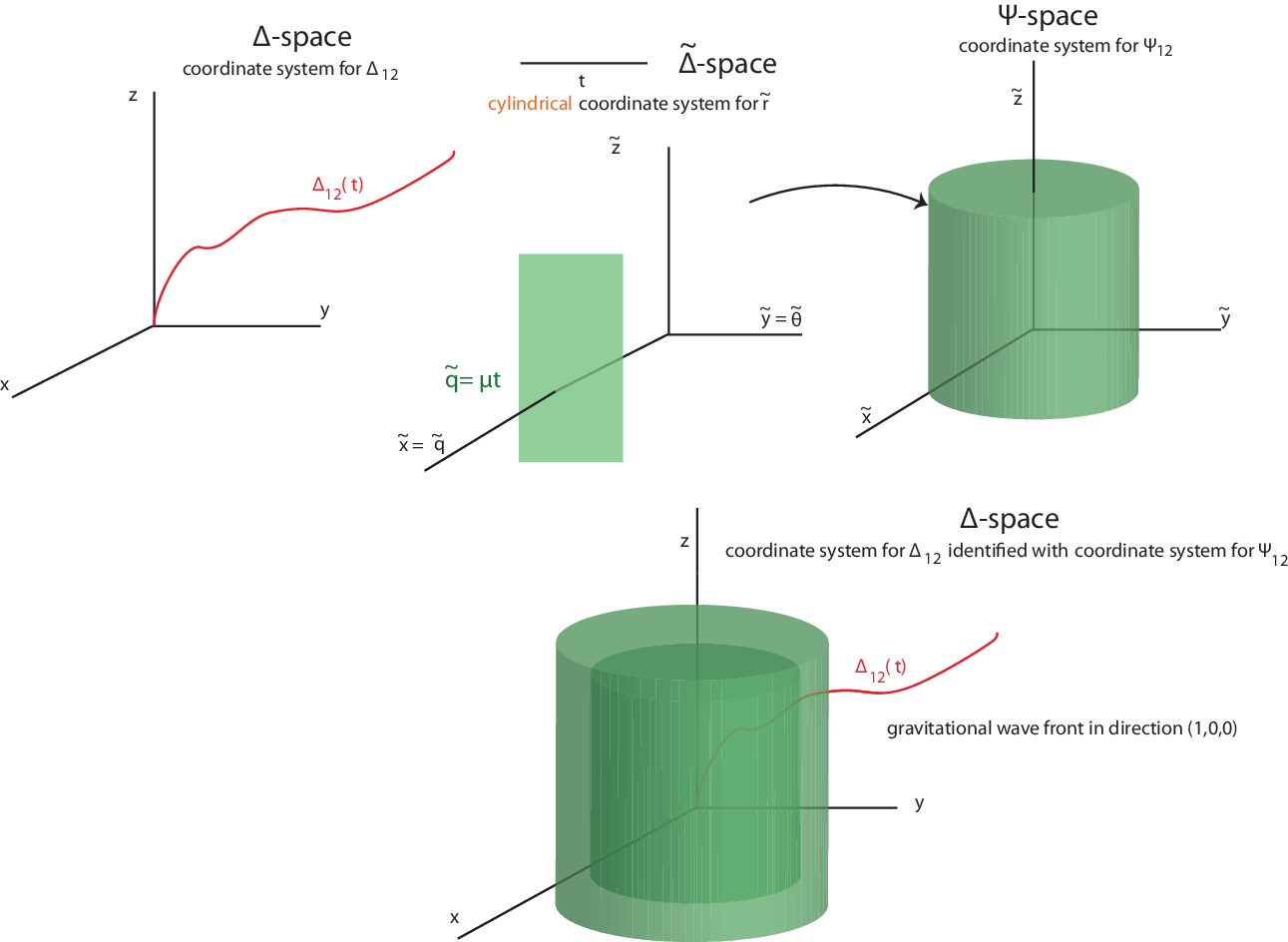}
\end{center}
\caption{Gravitational wave front in the direction $(1,0,0)$ resulting from cylindrical coordinates.}
\label{cylindricalwavefront}
\end{figure}

\medskip
We could also represent the $\tilde{\Delta}$-space in cylindrical coordinates $(\tilde{x} =\tilde{q},\tilde{y}=\tilde{\theta},\tilde{z})$ as shown in 
Figure \ref{cyclindricalcoordsys}.
As was the case for spherical coordinates, Definition
\ref{frontlocdef}, Proposition \ref{2bodyrelprop} are still valid, and the gradient of
Definition \ref{frontlocdef} is unbounded if $w =0$
However, since $w = v\cdot\tilde{r} -\mu t-c = v_1\tilde{q} + v_2\tilde{\theta} + v_3\tilde{z} -\mu t +c$, 
we find that $w = 0$ if and only if $v_1\tilde{q} + v_2\tilde{\theta} + v_3\tilde{z} = \mu t - c$.  As before,
we have the flexibility choose $v$, $c$ and $U = S_1/\|S_1\|$. Thus we set $c = 0$, and $u = v = (1,0,0)$.
Then $v_1\tilde{q} + v_2\tilde{\theta} + v_3\tilde{z} = \mu t - c$ simplifies to $\tilde{q} = \mu t$.
For fixed $\mu$ and fixed $t > 0$, the image of $\tilde{q}$ in the space associated with $\Psi_{12}(w)$ is a $z$-axis cylinder
with circular base $\tilde{x}^2 + \tilde{y}^2 = (\mu t)^2$, namely the set
$\{(\tilde{x},\tilde{y},\tilde{z})\mid \tilde{x}^2 + \tilde{y}^2 = (\mu t)^2\}$.
We then identify the $\Delta$-space with the $\Psi_{12}(w)$ space and note that
the gravitational wave front at time $t$ in the direction $(1,0,0)$ is the aforementioned cylinder. As $t$ increases,
the circular diameter of the cylinder increases to fill the $xy$-plane; see Figure \ref{cylindricalwavefront}.

\begin{figure}[ht]
\begin{center}
\includegraphics[height=1.9in,width=4.8in]{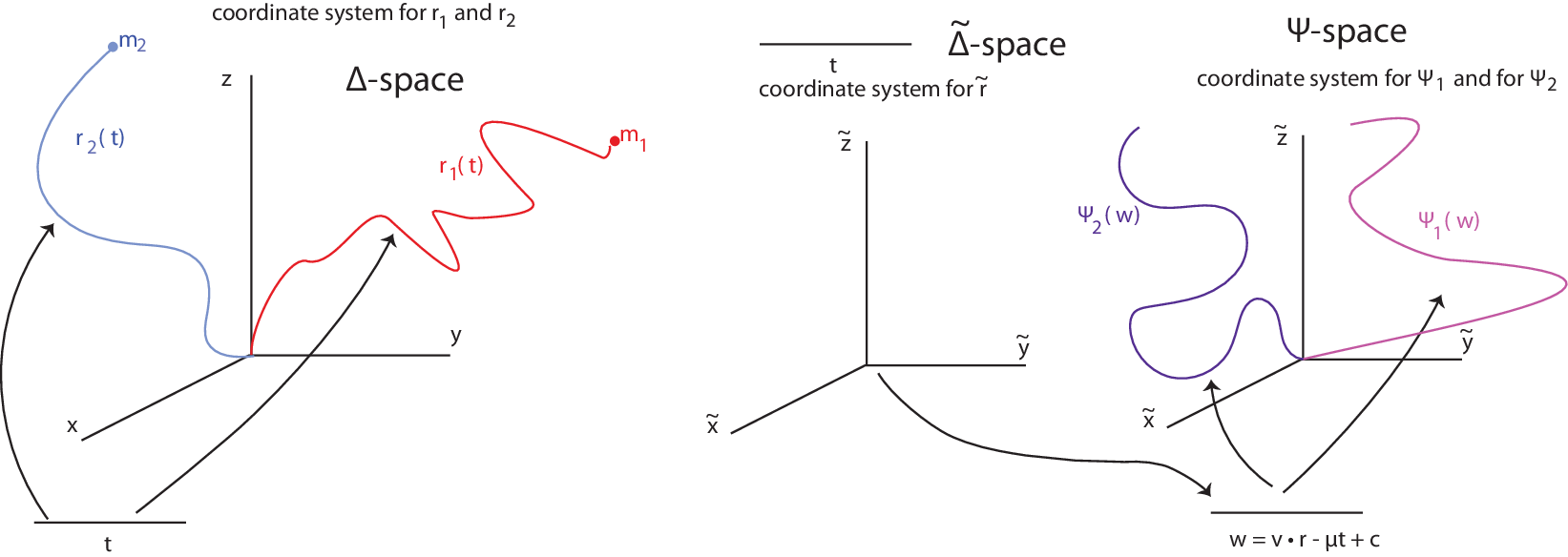}
\end{center}
\caption{Three copies of $\mathbb{R}^3$ for the $2$-body problem and its
companion wave equation.}
\label{adjustthreecorrspace}
\end{figure}

\medskip
For the wave fronts of the $2$-body companion wave equation, namely
\begin{align}\label{specialcaseN2vecformagain}
\left[\mu^2 - \lambda_1^2\|v\|^2\right]\frac{d^2\Psi_1}{dw^2} &= \frac{Gm_2(\Psi_2- \Psi_1)}{\|\Psi_2-\Psi_1\|^3}\notag\\
\left[\mu^2 - \lambda_2^2\|v\|^2\right]\frac{d^2\Psi_2}{dw^2} &= \frac{Gm_1(\Psi_1- \Psi_2)}{\|\Psi_2-\Psi_1\|^3},
\end{align}
we are still assuming that at $t =0$ and when $(\tilde{x},\tilde{y},\tilde{z}) = (0,0,0)$ in the $\Psi$-space, there is a collision between the
two point masses represented by $r_1(t)$ and $r_2(t)$, i.e. that $r_{1}(0) = (0,0,0) = r_2(0)$; see the discussion following
(\ref{newtonclosedformsolA}). Note that the $\Delta$-space contains the graphs of the curves $r_1(t)$ and $r_2(t)$, while the $\Psi$-space contains the 
graphs of the curves $\Psi_1(w)$ and $\Psi_2(w)$; see Figure \ref{adjustthreecorrspace}.
Definition \ref{frontlocdef} is adjusted as follows.
\begin{definition}\label{frontlocdefA}
 The front in the $\Delta$-space occurs when either of the gradients
 \[
 \|\left(\nabla_{(\tilde{x},\tilde{y},\tilde{z})}\Psi_{j1}(w),\partial \Psi_1/\partial t\right)\|,\quad
 \|\left(\nabla_{(\tilde{x},\tilde{y},\tilde{z})}\Psi_{j2}(w),\partial \Psi_2/\partial t\right)\|
 \]
  are undefined and/or unbounded for some $1\leq j\leq 3$. 
 \end{definition}
 Since Proposition \ref{Newton2bodyprop} implies that
 \[
 \Psi_{1}(w):= -\frac{\theta_1}{\theta_2}w^{2/3}S_1,\qquad \Psi_2 =^{2/3}S_1,
 \qquad S_1:= [s_{11}, s_{21}, s_{31}],
 \]
 where
 \[
 \frac{\theta_1}{\theta_2} = \frac{m_2[mu^2 -\lambda_2^2\|v\|^2]}{m_1[\mu^2 -\lambda_1^2\|v\|^2]}\neq 0,
 \]
 and since $w = v_1\tilde{x} + v_2\tilde{y} + v_3\tilde{z}-\mu t + c$, 
 we find that
 \begin{equation}\label{gradientvectagain1}
\left(\nabla_{(\tilde{x},\tilde{y},\tilde{z})}\Psi_{j1}(w),\frac{\partial \Psi_{j1}}{\partial t}\right)
=-\frac{\theta_1}{\theta_2}
\left(-\frac{2}{3}w^{-\frac{1}{3}}(v_1,v_2,v_3), -\frac{2}{3}w^{-\frac{1}{3}}\mu\right)s_{j1}
\end{equation}
and that
\begin{equation}\label{gradientvectagain2}
\left(\nabla_{(\tilde{x},\tilde{y},\tilde{z})}\Psi_{j2}(w),\frac{\partial \Psi_{j2}}{\partial t}\right)
=
\left(-\frac{2}{3}w^{-\frac{1}{3}}(v_1,v_2,v_3), -\frac{2}{3}w^{-\frac{1}{3}}\mu\right)s_{j1}.
\end{equation}
Hence, we once again conclude that (\ref{gradientvectagain1}) and (\ref{gradientvectagain2}) will be undefined when $w = 0$,
and we obtain the wave front which occurred for the relative $2$-body system.

\section*{Conflict of Interest}
The authors have no relevant financial or non-financial interests to disclose.
The authors have no conflicts of interest to declare that are relevant to the content of this article.

\section*{Data Sharing}
Data sharing is not applicable to this article as no new data were created or analyzed in this study.

 \end{document}